\documentclass[twocolumn, showpacs, superscriptaddress,preprintnumbers, amsmath, epsfig, floatfix, prl]{revtex4}

\usepackage{graphicx}
\usepackage{dcolumn}
\usepackage{bm}
\usepackage{ulem}

\usepackage{times}

\usepackage[utf8]{inputenc}
\usepackage[english]{babel}
\usepackage{keyval}
\graphicspath{{figs/}{figs:}{\figs}}
\setkeys{Gin}{width=0.90\columnwidth}

\usepackage[usenames,dvipsnames]{color}
\usepackage[]{ulem}

\newcommand{\comment}[1]{\textcolor{red}{#1}}
\renewcommand{\comment}[1]{\relax}

\newcommand{\todelete}[1]{\textcolor{green}{\sout{#1}}}
\renewcommand{\todelete}[1]{\relax}

\begin{document}
\title{Tetramer Orbital-Ordering induced Lattice-Chirality in Ferrimagnetic, Polar MnTi$_2$O$_4$}
\date{\today}
\author{A. Rahaman}
\affiliation{Department of Physics, Indian Institute of Technology Kharagpur, Kharagpur-721302, India}
\author{M. Chakraborty}
\affiliation{Centre for Theoretical Studies, Indian Institute of Technology Kharagpur, Kharagpur-721302, India}
\author{T. Paramanik}
\affiliation{Department of Physics, Indian Institute of Technology Kharagpur, Kharagpur-721302, India}
\affiliation{Department of Physics, School of Sciences, National Institute of Technology Andhra Pradesh, Tadepalligudem- 534102, India}
\author{R. K. Maurya}
\affiliation{School of Basic Sciences, Indian Institute of Technology Mandi-Kamand, Himachal Pradesh-175005, India}
\author{S. Mahana}
\affiliation{Rajdhani College, Bhubaneswar-751003, India}
\author{R. Bindu}
\affiliation{School of Basic Sciences, Indian Institute of Technology Mandi-Kamand, Himachal Pradesh-175005, India}
\author{D. Topwal}
\affiliation{Institute of Physics, Sachivalaya Marg, Bhubaneswar-751005, India}
\affiliation{Homi Bhabha National Institute, Training School Complex, Anushakti Nagar, Mumbai 400085, India}
\author{P. Mahadevan}
\affiliation{S. N. Bose National Center for Basic Sciences, Block JD, Salt Lake, Kolkata-700098, India}
\author{D. Choudhury}
\email{debraj@phy.iitkgp.ac.in}
\affiliation{Department of Physics, Indian Institute of Technology Kharagpur, Kharagpur-721302, India}

\begin{abstract}

\noindent Using density-functional theory calculations and experimental investigations on structural, magnetic and dielectric properties, we have elucidated
a unique tetragonal ground state for MnTi$_2$O$_4$, a Ti$^{3+}$ (3$\it{d}^1$)-ion containing spinel-oxide. With lowering of temperature around 164 K, cubic
MnTi$_2$O$_4$ undergoes a structural transition into a polar $\it{P}$4$_1$ tetragonal structure and at further lower temperatures, around 45 K, the system undergoes
a paramagnetic to ferrimagnetic transition. Magnetic superexchange interactions involving Mn and Ti spins and minimization of strain energy associated with co-operative Jahn-Teller distortions plays a critical role in stabilization of the unique tetramer-orbital ordered ground state which further gives rise to lattice chirality through subtle Ti-Ti bond-length modulations.

\end{abstract}

\maketitle

Transition-metal (TM) oxides with orbital degrees of freedom constitute a fascinating field of research and hosts copious physical phenomena, which include high-temperature superconductivity, colossal magnetoresistance and multiferroicity \cite{BKeimer2006,NNagaosa2000,SWCheong2007}. In transition metal oxides with strong electron-electron correlations, electrons are primarily localized on the atoms. Exotic physics ensue when such localized electrons also possess orbital degrees of freedom, i.e. electrons can choose to occupy between a set of equivalent and energy-degenerate atomic orbitals. Octahedrally coordinated Mn$^{3+}$ ions in LaMnO$_3$ with 3$\it{d}^4$ ($\it{t}_{\rm{2g}}^3$ - $\it{e}_{\rm{g}}^1$) configuration constitutes a representative example, where a single electron has a choice to occupy any of the two degenerate $\it{e}_{\rm{g}}$ orbitals. Often at a lower temperature, the electron chooses one from the two $\it{e}_{\rm{g}}$ orbitals, which breaks the local charge symmetry and is accompanied by differential oxygen-ion displacements, referred to as Jahn-Teller (JT) distortion. In a solid, such choices on different atoms are inter-dependent, which results into cooperative JT distortions associated with a spontaneous orbital-ordering (OO) transition, wherein localized occupied  orbitals on various ions form a regular pattern \cite{NNagaosa2000,YEndoh1998,YTokura1998,DDSarma2001}. Similar to the $\it{e}_{\rm{g}}$ OO systems, transition-metal oxides constituting ions possessing $\it{t}_{\rm{2g}}$-level orbital degrees of freedom, such as in YTiO$_3$ (one Ti$^{3+}$ $\it{d}$ electron in a subspace of three degenerate $\it{t}_{\rm{2g}}$ orbitals) \cite{MBibes2017,MImada2004}, MnV$_2$O$_4$ (two V$^{3+}$ $\it{d}$ electrons among three degenerate $\it{t}_{\rm{2g}}$ orbitals) \cite{SENagler2008,TKatsufuji2007,TSahaDasgupta2009}, also exhibit cooperative JT distortions and various OO ground states.
Mostly, in these TM systems, either a ferro-OO state (similar occupied orbital at all ionic sites) or an antiferro-OO state (with alternate ions occupied by similar orbitals), or a combination of the two along different directions is realized.
The presence of higher-order OO has very few examples, such as CuIr$_2$S$_4$ \cite{SWCheong2002,TMizokawa2005} and Fe$_3$O$_4$ \cite{DJHuang2004,JPAttfield2012,Wright2001}, and unlike the simpler examples discussed earlier, the forces driving the OO still remain a puzzle.


\indent In this letter, we report a unique tetramer OO state in spinel oxide MnTi$_2$O$_4$ (which contains octahedrally-coordinated Ti$^{3+}$-3$\it{d}^1$ ions). As this is unusual, we use a combination of theory and experiments to explore the driving mechanism for the orbital ordering. The ground-state lattice and magnetic structure of MnTi$_2$O$_4$, however, remains contentious \cite{Sonehara,Huang,Zhang}. We elucidate that the ferrimagnetic tetragonal $\it{P}$4$_1$ structure is the ground-state of MnTi$_2$O$_4$ and show that this structure hosts a unique combination of tetramer OO, lattice-chiralilty and spontaneous electric polarization. As the levels in one spin channel on Mn are filled, superexchange interactions between Mn and Ti sites results in an antiferromagnetic Mn-Ti coupling. This in turn leads to a ferromagnetic coupling between the spins on Ti.  Ti$^{3+}$ ions are JT active and so while one can envisage few patterns of orbital ordering consistent with a ferromagnetic Ti lattice, in this system we find that the strain energy costs are lowest when a tetramer ordering is favoured. Thus, for the first time, not only do we identify an unusual orbital ordering in MnTi$_2$O$_4$, we also identify the microscopic considerations that drive it.

\begin{figure}[h]
\vspace*{-0.00 in}
\hspace*{-0.12 in}\scalebox{1.15}
{\includegraphics{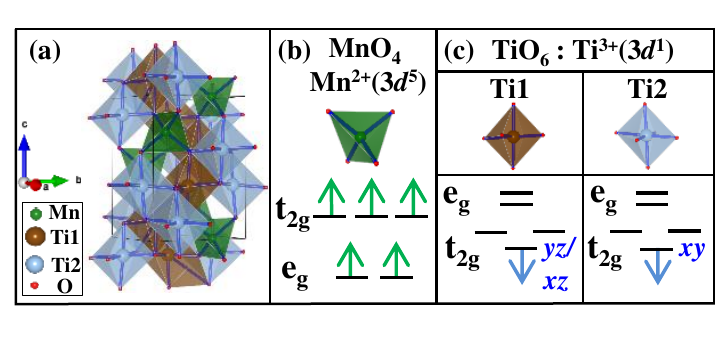}}
\vspace*{-0.35 in}\caption{(color online) (a) Schematic of the tetragonal $\it{P}$4$_1$ spinel structure of MnTi$_2$O$_4$ with MnO$_4$ tetrahedral units and two-kinds of TiO$_6$ octahedral units. Ground state orbital and spin configurations of (b) Mn$^{2+}$ and (c) Ti$^{3+}$ (Ti1 and Ti2) ions. The single Ti 3$\it{d}^{\rm1}$ electron occupies either the $|\it{xz}>$ or the $|\it{yz}>$ orbital for the Ti1 site and the $|\it{xy}>$ orbital for the Ti2 site.}\label{Structure}
\end{figure}

Ab-initio density-functional theory (DFT) calculations were performed using all electron full-potential augmented linearized plane wave method taking augmented plane wave basis as implemented in WIEN2k code \cite{Blaha}.
In order to elucidate the ground state of MnTi$_2$O$_4$, relative energies between various spinel structures were investigated and spin-polarized calculations with different spin configurations were performed. Volume as well as internal geometries were optimized for all the investigated structures in presence of on-site electron-electron correlation ($\it{U}$) using GGA-PBE exchange correlation functional \cite{Perdew}. Spin-orbit coupling was incorporated with GGA+U calculations by second variational code along with scalar relativistic functions \cite{SO}, however, its effect was found to be negligible. Muffin-tin radius of 1.98, 1.80 and 1.50 a.u. for Mn, Ti and O, respectively and a k-point mesh of 9{$\times$}9{$\times$}6 were considered for all the calculations. Rkmax, Gmax and lmax were set to 7.0, 14.0 Bohr$^{-1}$ and 12, respectively. Throughout the manuscript, the Ti orbitals are defined in the local TiO$_6$ coordinate system. The ferroelectric polarization calculations were performed using Berry-phase method \cite{DVanderbilt1993,RResta1994} with Vienna Ab initio Simulation Package (VASP) \cite{GKresse1993}. For the experimental investigations, polycrystalline sample of MnTi$_2$O$_4$ was synthesized using solid state reaction route from a mixture of MnO, TiO$_2$ and metallic Ti powders \cite{Huang}. The mixture was ground well and sintered in the form of pellets at 900$^0$C under vacuum in a sealed quartz tube. The phase formation of the sample, which contained $\sim$5$\%$ of Ti$_2$O$_3$ impurity phase (as reported earlier \cite{Sonehara}), was established using temperature dependent x-ray diffraction (XRD) technique and cell-parameters were extracted from Rietveld refinement of XRD data using FULLPROF package \cite{Fullproof}. X-ray absorption near-edge structure (XANES) and Extended x-ray absorption fine-structure (EXAFS) measurements were carried out to investigate the electronic and local-crystallographic structures at various temperatures at P-65 beamline at PETRA III synchrotron source, DESY, Hamburg, Germany. The pre-processing and fitting of EXAFS data were carried out over the $\it{k}$-range of 3-12 $\mathring{\rm{A}}^{-1}$ using ATHENA and ARTEMIS softwares \cite{Athena}.

\indent Cubic $\it{F}$d-3m as well as tetragonal $\it{I}$4$_1$/amd (which is a simple elongation of cubic $\it{F}$d-3m along the $\it{c}$-axis) structures have been proposed to be the ground state of MnTi$_2$O$_4$ \cite{Sonehara,Huang,Zhang}. We, however, find that the ferrimagnetic $\it{P}$4$_1$ tetragonal structure possesses the lowest energy (56 meV/f.u. smaller than the closest cubic $\it{F}$d-3m structure). Between the cubic $\it{F}$d-3m structure and tetragonal $\it{I}$4$_1$/amd structure of MnTi$_2$O$_4$, $\it{F}$d-3m structure is found to have lower energy, which is in consistence with the earlier report \cite{Zhang}. The obtained ground-state $\it{P}$4$_1$ structure of MnTi$_2$O$_4$ consists of two inequivalent Ti sites (as shown in Fig.\ref{Structure}(a)), and, is, thus, lower in symmetry from the $\it{I}$4$_1$/amd tetragonal structure, which has only one Ti site. Interestingly, the $\it{P}$4$_1$ structure is non-centrosymmetric, with both Ti and Mn ions displaced from the centre of the TiO$_6$ octahedral and MnO$_4$ tetrahedral cages, respectively. The calculated ferroelectric polarization value of the $\it{P}$4$_1$ structure of MnTi$_2$O$_4$ is 0.5 $\mu$C cm$^{-2}$.


\begin{figure}[h]
\scalebox{1.3}
{\includegraphics{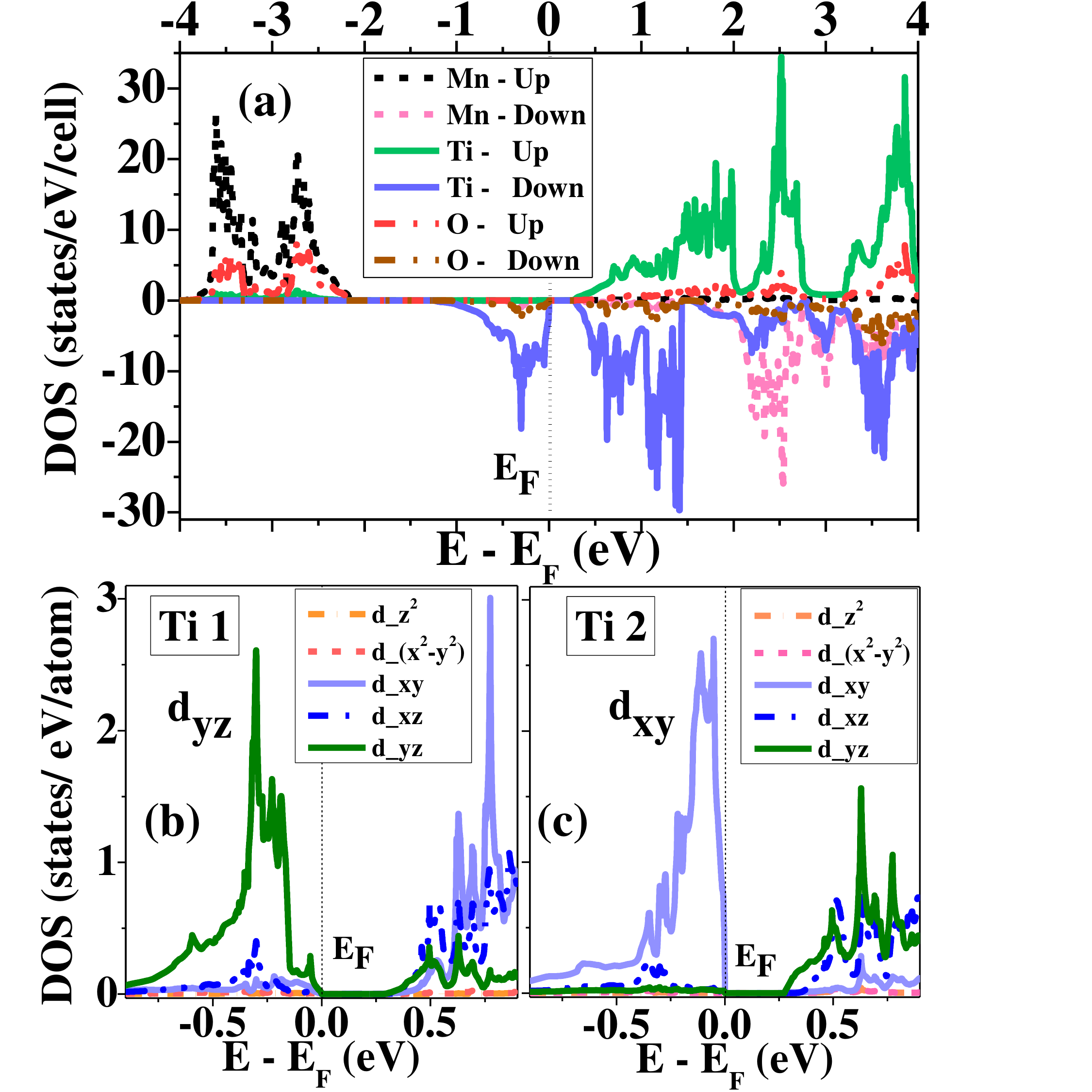}}
\vspace*{-0.1 in}\caption{(color online) Density-of-states (DOS) of $\it{P}$4$_1$ structure of MnTi$_2$O$_4$. (a) shows total-DOS for all atoms (Mn, Ti, O), (b) and (c) show partial-DOS for Ti1 $\it{d}$-levels and Ti2 $\it{d}$-levels, respectively.}\label{DOS}
\end{figure}

\begin{figure}[h]
\hspace*{-0.05 in}\scalebox{1.15}
{\includegraphics{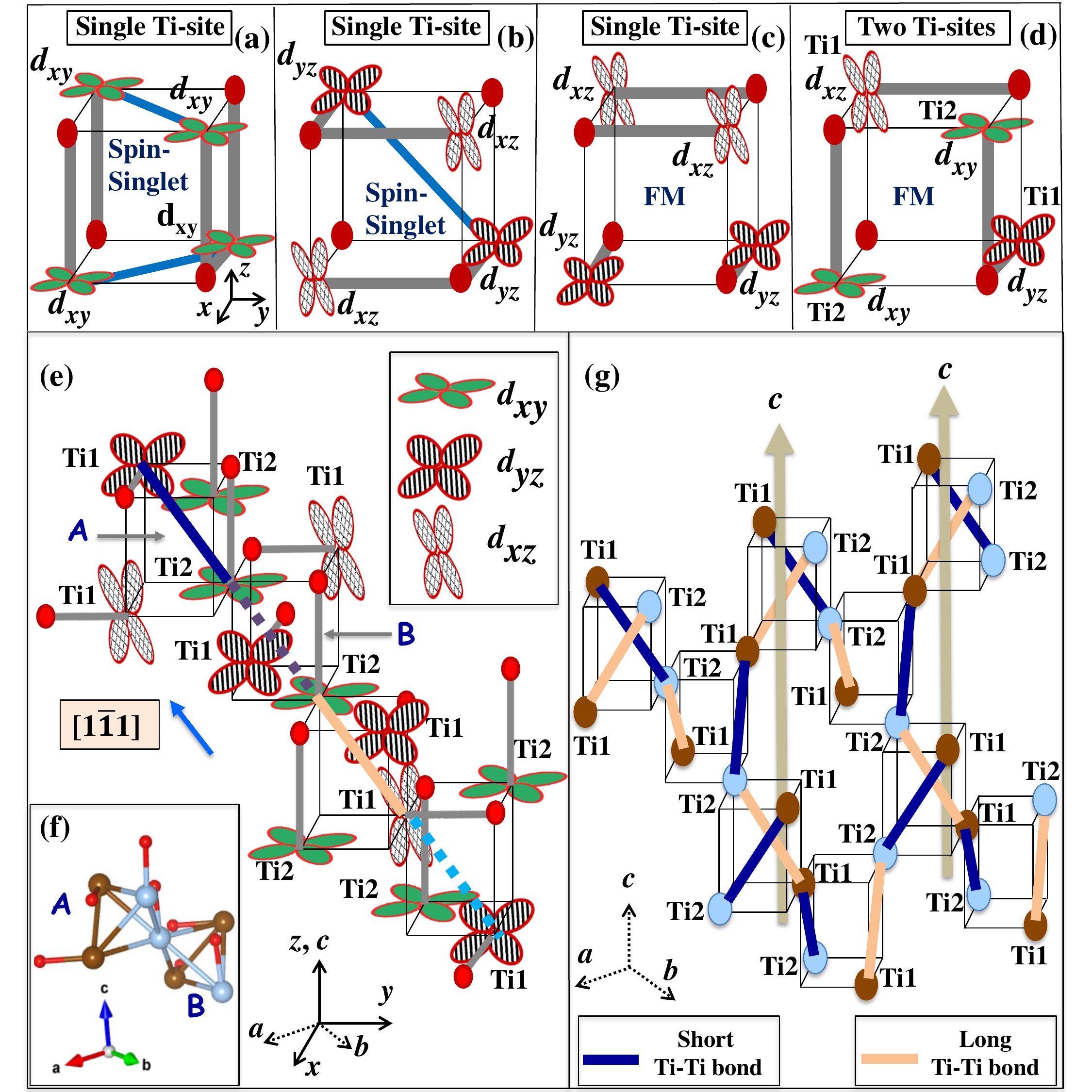}}
\vspace*{-0.04 in}\caption{(color online) Illustration showing various possible distributions of Ti-site orbital occupancies among the Ti-tetrahedra resulting from Jahn-Teller distortions for ((a)-(c)) a single kind of Ti-atom and (d) two kinds of Ti-atoms in the spinel structure. For each Ti-atom, the direction of the shortest Ti-O bond is illustrated by thick grey lines. Small red balls are the O atoms. Direct overlap of similar occupied orbitals in (a) and (b) would result in formation of spin-singlet dimers, which are highlighted by thick blue lines. (e) Illustration showing the Ti $\it{d}$-level - $\it{d}_{\it{yz}}$-$\it{d}_{\it{xz}}$-$\it{d}_{\rm{xy}}$-$\it{d}_{\rm{xy}}$ - tetramer orbital ordering along [1$\bar{\rm{1}}$1] direction of the crystal unit-cell of MnTi$_2$O$_4$. The obtained Ti $\it{d}$ orbital-ordering is accompanied with Ti-Ti bond-length modulations (the solid dark blue, dashed blue, solid light-brown and dashed sky blue colored Ti-Ti bond lengths are 3.010 $\mathring{\rm{A}}$, 3.129 $\mathring{\rm{A}}$, 3.145 $\mathring{\rm{A}}$, and 3.025 $\mathring{\rm{A}}$, respectively.) (f) The spatial-distribution of the shortest Ti – O bond distances in the corner-shared Ti tetrahedral network of MnTi$_2$O$_4$. (g) Short and long Ti-Ti bonds, when joined, form helices around the crystallographic $\it{c}$-axis of MnTi$_2$O$_4$.}\label{OO}
\end{figure}

Calculations performed using different U values, like (2,2), (3,2) and (3,3) eV for (Mn,Ti) ions (U values taken in accordance with earlier calculations on related systems \cite{LCraco2008,TSahaDasgupta2009,ATaraphder2016}), give similar DOS and same OO pattern and differ in their band-gap values. Fig.\ref{DOS}(a) shows the spin-resolved total density of states (DOS) for the ground-state $\it{P}$4$_1$ structure of MnTi$_2$O$_4$, evaluated using GGA+U calculation ($\it{U}$=(3,2) eV for (Mn,Ti) ions). The total DOS near the fermi level is dominated by Ti DOS and its splitting gives rise to an insulating band-gap of 0.26 eV. As seen in Fig.\ref{DOS}(a), the up-spin channel is nearly completely occupied for the Mn atoms, validating its 2+ (3$\it{d}^5$) high-spin state. The up-spin channel for Ti atoms is nearly empty, signifying the ferrimagnetic configuration where Mn (with only up-spin levels occupied) and Ti spins are aligned antiferromagnetically (shown in Fig.\ref{Structure}(b)) and the Ti spins arranged ferromagnetically. The near-orthogonal superexchange interactions between Mn-Ti ions (average Mn$^{2+}$-O$^{2-}$-Ti$^{3+}$ bond angle is $\sim\rm{123}^o$) and Ti-Ti ions (average Ti$^{3+}$-O$^{2-}$-Ti$^{3+}$ bond angle is $\sim\rm{95.9}^o$) in MnTi$_2$O$_4$, following Goodenough-Kanamori-Anderson rules \cite{DIKhomskii2014}, is also in accordance with the obtained ferrimagnetic coupling between Mn$^{2+}$ (with only one hopping-spin channel available) and Ti$^{3+}$ ions and a ferromagnetic Ti$^{3+}$ lattice. Further, Fig.\ref{DOS}(b) and (c) show the orbitally-resolved partial DOS for the two inequivalent Ti atoms, Ti1 and Ti2. Clearly, the $\it{t}_{2\rm{g}}$-level degeneracy of the TiO$_6$ octahedra of the cubic state is broken in the tetragonal phase through JT distortions and the single Ti$^{3+}$-3$\it{d}^1$ electron dominantly occupies a single $\it{t}_{2\rm{g}}$-$\it{d}$ level, which varies between the Ti atoms. The occupied $\it{t}_{2\rm{g}}$ orbital is one among the a{$_{1\rm{g}}$} (d$_{xz}$ or d$_{yz}$) orbitals for the Ti1 atoms and the e${_g}'$ (d$_{xy}$) orbital for the Ti2 atoms. The $\it{d}_{\it{xz}}$, $\it{d}_{\rm{yz}}$ and $\it{d}_{\rm{xy}}$ orbital occupancies are associated with the shortest Ti-O bond being along the $\it{y}$, $\it{x}$, and $\it{z}$ ($\it{c}$) directions, respectively. Instead of two inequivalent Ti atoms, presence of a single kind of Ti atom would have either lead to a spin-singlet pairing (indicated by diagonal blue lines) among the Ti spins \cite{DiMatteo2004,Schmidt2004,YUeda2002}, as indicated in Fig.\ref{OO}(a) or (b), or resulted in unsustainable piling of strain along certain crystallographic directions from accompanying JT effects, as indicated in Fig.\ref{OO}(c). Presence of two distinct Ti sites, as shown in Fig.\ref{OO}(d), while ensures a ferromagnetic coupling between the Ti$^{3+}$ ions, also results in effective distribution of the shortest Ti-O bonds amongst different crystallographic directions, thereby leading to a reduction of the cooperative JT-effect related strain energy, as shown in Fig.\ref{OO}(d) and (f).



The obtained charge-density plots of the occupied Ti $\it{d}$-orbitals illustrates the emergence of OO in the ground state involving all three $\it{t}_{2\rm{g}}$ orbitals ($\it{d}_{\it{xz}}$, $\it{d}_{\rm{yz}}$ and $\it{d}_{\rm{xy}}$). In subsequent $\it{ab}$-planes, the OO pattern remains the same, only the character of the participating orbitals vary between either the $\it{d}_{\it{xz}}$-$\it{d}_{\rm{xy}}$ or the $\it{d}_{\rm{yz}}$-$\it{d}_{\rm{xy}}$ pairs. Interestingly, a unique tetramer-OO, i.e. a $\it{d}_{\it{yz}}$-$\it{d}_{\it{xz}}$-$\it{d}_{\rm{xy}}$-$\it{d}_{\rm{xy}}$ ordering is observed for the Ti-chains running along the equivalent $<$111$>$ directions. The tetrahedrons comprising two distinct Ti sites, as shown in Fig.\ref{OO}(d), form interconnected chains and a unique tetramer atomic-ordering (Ti1-Ti1-Ti2-Ti2) along the equivalent $<$111$>$ directions (as shown in Fig.\ref{OO}(e)). Similar consideration of distribution of shortest Ti-O bonds amongst three orthogonal directions and reduction of JT-effect related strain energy, as effective in giving rise to two distinct Ti-sites in Fig.\ref{OO}(d), also becomes effective for the interconnected Ti chains along equivalent $<$111$>$ directions, resulting in a unique tetramer-OO ($\it{d}_{\it{yz}}$-$\it{d}_{\it{xz}}$-$\it{d}_{\rm{xy}}$-$\it{d}_{\rm{xy}}$) state, as shown in Fig.\ref{OO}(e). Modulations in Ti-Ti bond distances are necessarily associated with such cooperative JT distortions and we identify four Ti-Ti bond distances in the tetramer OO state (shown in Fig. \ref{OO}(e)). Interestingly, among these four Ti -Ti bonds, the short and the long Ti-Ti bonds, when joined together, form helices with a particular winding direction along the crystallographic $\it{c}$ direction, causing the structure to become lattice chiral (illustrated in Fig. \ref{OO}(g)). We also find that a tetragonal $\it{P}$4$_3$ structure is degenerate in energy to the $\it{P}$4$_1$ structure, and these two structures vary in the sense of the lattice chirality (one is left-handed chiral and the other is right-handed chiral) and are similar otherwise.

\begin{figure}[t]
\vspace*{-0.09 in}
\hspace*{-0.09 in}\scalebox{1.15}
{\includegraphics {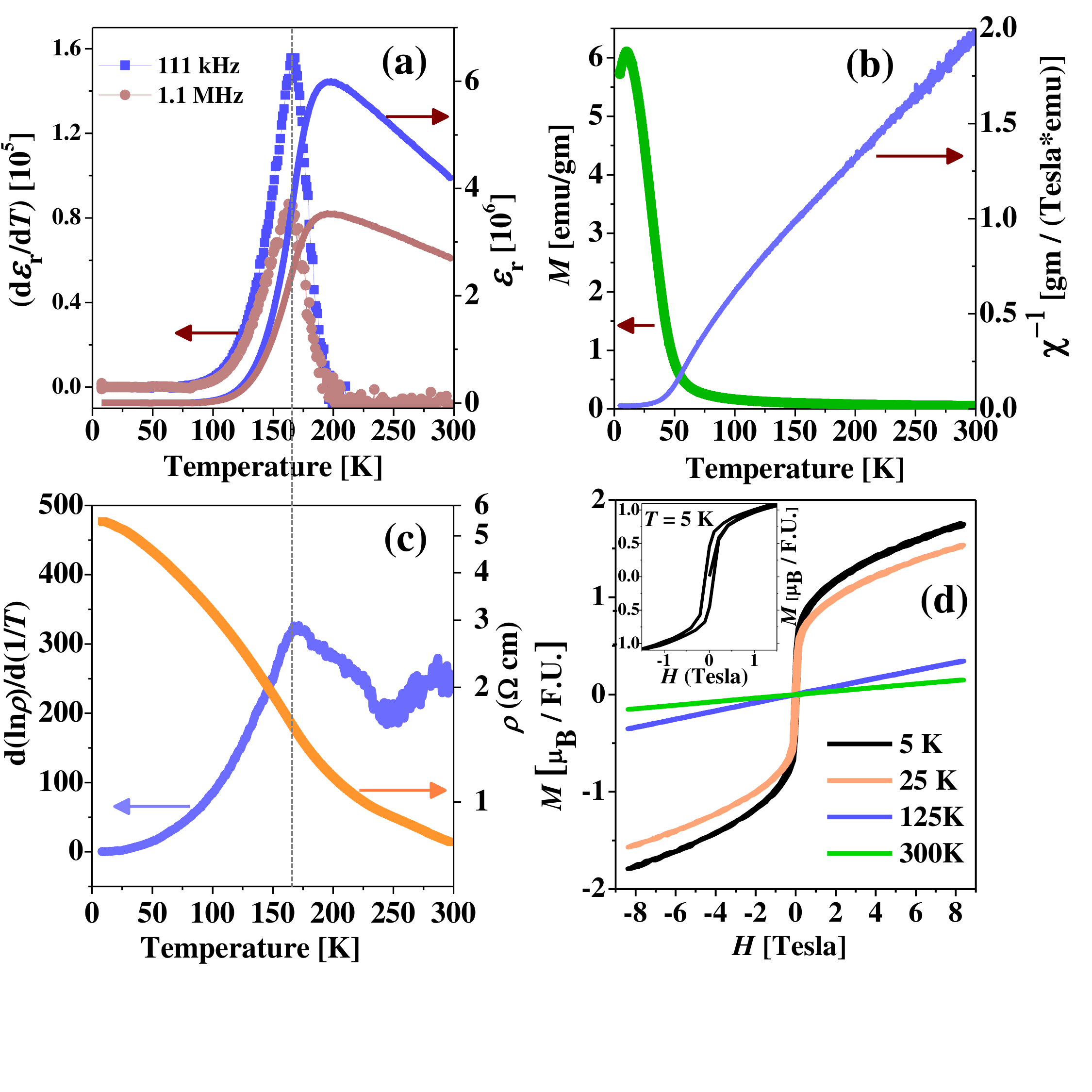}}
\vspace*{-0.53 in}\caption{(color online) Temperature ($\it{T}$)- dependencies of (a) dielectric constant ($\varepsilon_{\rm{r}}$) and $\frac{\rm{d}\varepsilon_{\rm{r}}}{\rm{d}T}$, (b) magnetization ($\it{M}$) and inverse-susceptibility ($\frac{1}{\chi}$) measured with a magnetic-field ($\it{H}$) of 0.1 Tesla, (c) resistivity ($\it{\rho}$) and $\frac{\rm{dln}\rho}{\rm{d}(1/T)}$ of MnTi$_2$O$_4$. (d) Isothermal $\it{M}$-$\it{H}$ curves measured at 5 K, 25 K, 125 K and 300 K with the inset showing an expanded view of the hysteresis loop at 5 K.}\label{RMT}
\end{figure}

\begin{figure}[h]
\scalebox{1.4}
{\includegraphics{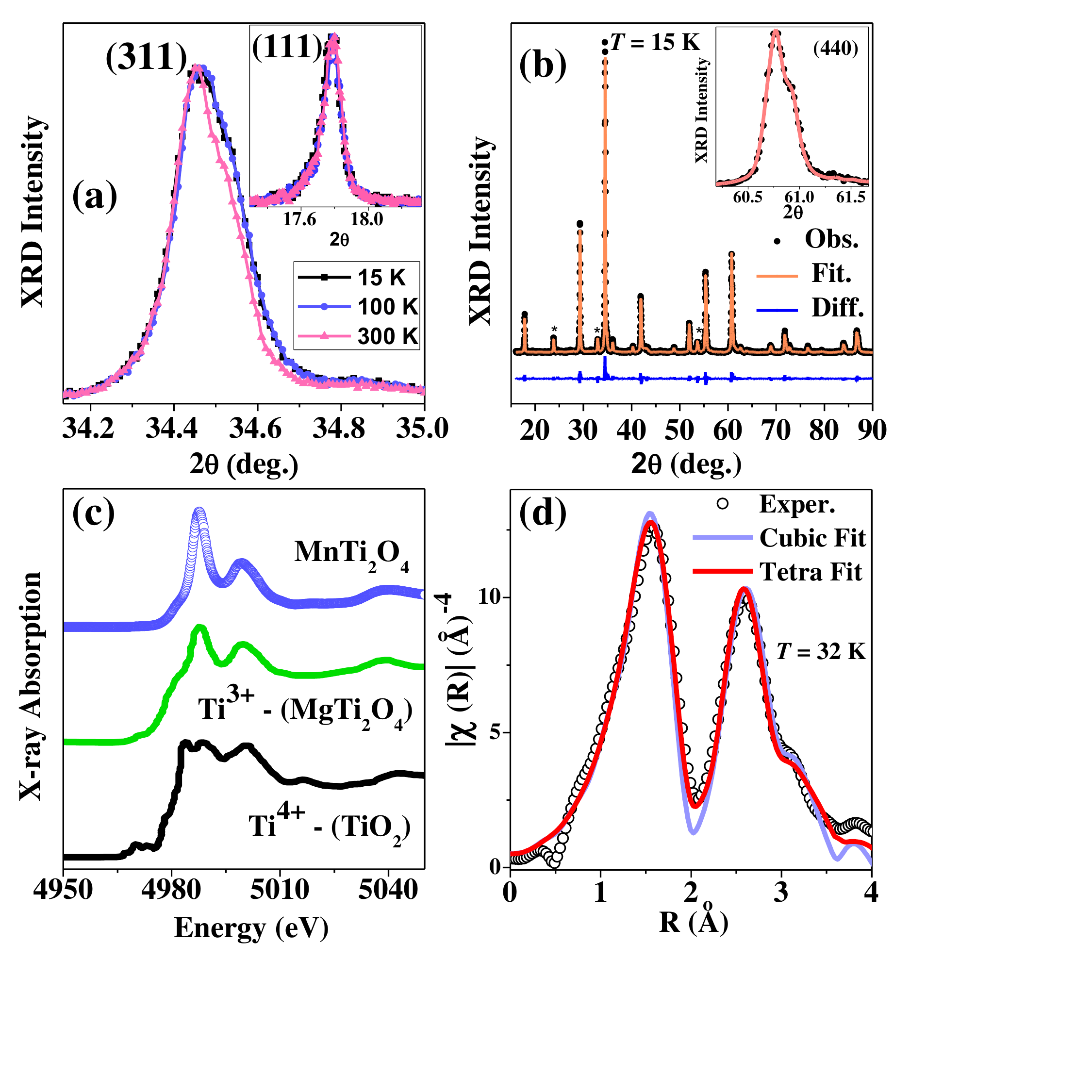}}
\vspace*{-0.7 in}\caption{(color online) (a) Broadening of (311) x-ray diffraction (XRD) peaks measured at $\it{T}$ = 15 K and 100 K with respect to that measured at $\it{T}$ = 300 K. Inset shows that similar $\it{T}$-dependent broadening is not observed for (111) XRD peak. (b) Fitting of XRD spectrum at $\it{T}$ = 15 K with a $\it{P}$4$_1$ tetragonal structure. The refinement also includes $\sim$5$\%$ Ti$_2$O$_3$ impurity phase, whose peaks are marked with asterisks. (c) Comparison of Ti-$\it{K}$-edge XANES spectra of MnTi$_2$O$_4$ with MgTi$_2$O$_4$ (which contains Ti$^{3+}$ ions) \cite{MgTi2O4} and TiO$_2$ (which contains Ti$^{4+}$ ions) \cite{TiO2}. (d) Comparative fittings of Ti-$\it{K}$-edge EXAFS oscillations at $\it{T}$ = 32 K with a tetragonal and a cubic structure.}\label{ExpStrct}
\end{figure}

To corroborate the theoretical findings of a structural transition, we next discuss the results of experimental investigations on MnTi$_2$O$_4$. The temperature-dependence of dielectric constant of MnTi$_2$O$_4$ is plotted in Fig.\ref{RMT}(a). The dielectric constant rises sharply from low-temperatures and its derivative exhibits a clear peak at $\sim$164 K. Importantly, this peak position does not disperse with varying electric-field frequencies, which is indicative of a ferrolectric transition. A clear transition can also be easily discerned at $\sim$164 K in the plots of the temperature-dependencies of resistivity and scaled effective activation energy (in units of K), as shown in Fig.\ref{RMT}(c). Temperature-dependent XRD studies were carried out to investigate the presence of a structural transition around this temperature range. Differential XRD peak-broadenings were indeed observed for MnTi$_2$O$_4$ with lowering of temperature. For some representative XRD peaks (the characteristic right-shoulder arises from Cu $\it{k}_\beta$ satelite), like (311), (400), (511) and (440), the peaks were found to be broader at lower temperatures (15 and 100 K) compared to the 300 K spectrum (the former comparison shown in Fig. \ref{ExpStrct}(a)). We note that this behavior is opposite to what is expected from usual thermal-broadening effect. The (111) XRD peak of MnTi$_2$O$_4$, as shown in the inset to Fig. \ref{ExpStrct}(a), however, exhibits no additional broadening with lowering of temperature. The observed differential XRD peak-broadenings is in consistence with a cubic to low-temperature tetragonal structural transition and has been used as a characteristic tool to identify temperature-dependent structural transition \cite{Sonehara}. Rietveld-refinements of the XRD pattern of MnTi$_2$O$_4$ at 15 K were performed considering various structures, like cubic $\it{F}$d-3m, orthorhombic $\it{F}$ddd, tetragonal $\it{I}$4$_1$/amd and tetragonal $\it{P}$4$_1$. We find that the best refinement for the low-temperature ($\it{T}$=15 K) XRD spectrum (shown in Fig. \ref{ExpStrct}(b)) is obtained using the tetragonal non-centrosymmetric enantiomorphic $\it{P}$4$_1$ structure ($\chi^2$ of 1.50). To investigate the low-$\it{T}$ structural distortions in further details, we have carried out temperature-dependent Ti-$\it{K}$- edge XANES and EXAFS spectroscopic studies on MnTi$_2$O$_4$. First, we discuss the spectral shape of the Ti-$\it{K}$-edge XANES spectrum which is a bulk-sensitive probe for the valence state of Ti ions. The comparisons of Ti-$\it{K}$-edge spectral shape of MnTi$_2$O$_4$ with standard reference spectra, as shown in Fig. \ref{ExpStrct}(c), establish the presence of Ti ions in the (3+) valence-state in nearly-stoichiometric MnTi$_2$O$_4$, in consistence with the estimated valence state. We next discuss the corresponding EXAFS oscillations, which is an excellent probe of local structural distortions, complimentary to the bulk-sensitive XRD technique. In the following, we focus on analyses of EXAFS oscillations at Ti-$\it{K}$-edge recorded on either sides of the estimated structural transition temperature of $\sim$164 K, i.e. 32 K and 200 K. The observed EXAFS oscillations at 32 K, as shown in Fig.\ref{ExpStrct}(d), were fit better (both qualitatively in terms of matching spectral shape and quantitatively in terms of a lower fitting-related $\it{R}$ factor) using a tetragonal structure than a cubic structure. Also, at 200 K, the EXAFS oscillations were fit better with the cubic phase than the tetragonal structure. Thus, both bulk-sensitive XRD and local-probe EXAFS investigations compliment each other well and suggest a low-temperature $\it{P}$4$_1$ tetragonal ground-state structure for MnTi$_2$O$_4$, in accordance with our theoretical results.

Magnetization ($\it{M}$) of MnTi$_2$O$_4$ rises sharply on lowering of temperature ($\it{T}$) around 50 K, as shown in Fig.\ref{RMT}(b). A magnetic transition around 45 K is estimated from the peak position in the deriavative of $\it{M}$-$\it{T}$ data (the small impurity phase of Ti$_2$O$_3$, which has a spin-singlet transition at $\sim$450 K \cite{LHTjeng2018}, does not affect the magnetic data in the investigated temperature range), which is in consistence with corresponsing ESR data \cite{Huang}. A negative value for the temperature intercept, i.e. -42 K, is also obtained from a Curie-Weiss analysis of the inverse magnetic susceptibility data (shown in Fig.\ref{RMT}(b)). Further, observation of clear $\it{M}$ vs. magnetic field ($\it{H}$) loops below 45 K (shown in inset of Fig.\ref{RMT}(d)), establish the presence of a ferrimagnetic transition at 45 K for MnTi$_2$O$_4$, which is in excellent agreement with our theoretical results.


In summary, using ab-initio DFT calculations and a combination of several experimental techniques we have elucidated a unique $\it{P}$4$_1$ tetragonal ground state of MnTi$_2$O$_4$, which hosts a unique combination of tetramer OO, ferroelectricity and lattice-chirality. The obtained Ti-site spin and orbital configurations are in good agreement with results from model calculations involving spin-orbital-superexchange interactions of three-fold orbitally degenerate $\it{S}$ = $\frac{1}{2}$ ions on a general pyrochlore lattice ($\it{B}$-lattice in $\it{AB}_2O_4$ spinels) \cite{DiMatteo2004}. We find that a combination of SE interactions among Mn and Ti spins and consideration of minimization of cooperative JT effect related strain energy becomes instrumental in stabilization of the unique ground state in MnTi$_2$O$_4$.

A.R. performed the theoretical calculations, partly using VASP, for which D.C. and A.R. would like to thank Swastika Chatterjee for her support. D.C. would like to gratefully acknowledge SRIC-IIT Kharagpur (ISIRD grant), SERB, DST (funding under project file no. ECR/2016/000019) and BRNS, DAE (funding through sanction number 37(3)/20/23/2016-BRNS) for financial support. T.P. would like to acknowledge SERB for providing fellowship (file no. PDF/2016/002580). S.M. and D.T. would like to gratefully acknowledge financial support by DST provided with in the framework of the India@DESY collaboration. A.R. and D.C. would like to acknowledge Poonam Kumari, Partha Pratim Jana, Arghya Taraphder and Dibyendu Dey for various fruitful discussions.

\end{document}